\documentclass{epl}

\newcommand{\drp}[2]{\frac{\partial #1}{\partial #2}}

\title{Viscous force exerted on a foam at a solid boundary : \\influence of the liquid fraction and of the bubble size.}
\author{E. Terriac\inst{1} \and J. Etrillard\inst{1} \and  I. Cantat\inst{1}}
\institute{
  \inst{1} GMCM UMR 6626, Universit\'e Rennes I, 35042 Rennes cedex, France
}
\pacs{83.50.Ha}{Flow in channels}
\pacs{83.80.Iz}{Emulsions and foams}
\pacs{83.85.Jn}{Viscosity measurements}

\begin{document}

\maketitle

\begin{abstract} We study experimentally the pressure drop needed to push a bubble train in a millimetric channel, as a function of the velocity. For dry liquid foams, the influence of the amount of liquid and of the bubble size is pointed out and we predict theoretically that this influence is closely related to the power law obtained for the force/velocity relation. This model is in fair agreement with our experimental data and provides a new interpretation of previous results.
\end{abstract}

\section{Introduction}
The study of the dissipation induced by foam flows in confined geometries was initially mainly motivated by the petroleum industry  and the need to push 
multiphasic liquids through porous rocks\cite{hirasaki85,chambers90}. The domains of application recently enlarged to very different fields as microfluidics \cite{cubaud04}, 2D foams in Hele Shaw cells, seen as model systems for complex fluids \cite{kern04,dollet05,cantat03a}, or even lung surfactant 
foam flows in pulmonary airways \cite{howell00}, thus giving a revived interest to this old problem. 
When a soap film train is pushed in a channel by a constant gas flux, 
the pressure drop may be several orders 
of magnitude larger than the one needed to push only water, at the same flux. 
Moreover, if the amount of gas in the bubble train increases, reaching more than $95 \%$ in volume, the pressure 
drop even increases. This puzzling multiphase flow behavior is governed by low Reynolds number hydrodynamical processes, with 
subtle non linearities related to the deformable air/liquid interfaces. 
Despite a large number of theoretical  and experimental studies, for isolated bubbles \cite{bretherton61,wong95,schwartz86} or bubble trains \cite{ratulowski89,xu03,cantat04,denkov05}, a coherent 
understanding of the phenomenon over the whole parameter range explored is still missing. Analytical approaches, mainly based on the Landau Levich lubrication theory \cite{landau42} predict a pressure drop scaling as the bubble velocity at the power $2/3$ \cite{bretherton61}. Nevertheless, 
as already mentioned by Schwartz {\it et al} \cite{schwartz86}, a phenomenological $1/2$ power law very often better agrees with experimental 
observations. In a recent paper \cite{denkov05}, these two behaviors were experimentally observed. The change between the $1/2$ and the $2/3$ power laws was obtained by changing the surfactant monolayer rigidity, a parameter neglected in the previous theoretical treatments.
The aim of this paper is to determine experimentally, for dry foams, the relation between the foam velocity, its  texture (liquid fraction and bubble size) and the viscous force exerted on the channel wall. Once rescaled by an adimensional number involving a non trivial combination of these parameters, all the experimental data are superimposed on a single master curve. 
This rescaling, obtained by lubrication analysis, provides a new interpretation of previous results.

\section{Experimental set-up}

\begin{figure}
\begin{minipage}{7cm}
\includegraphics[width=6cm]{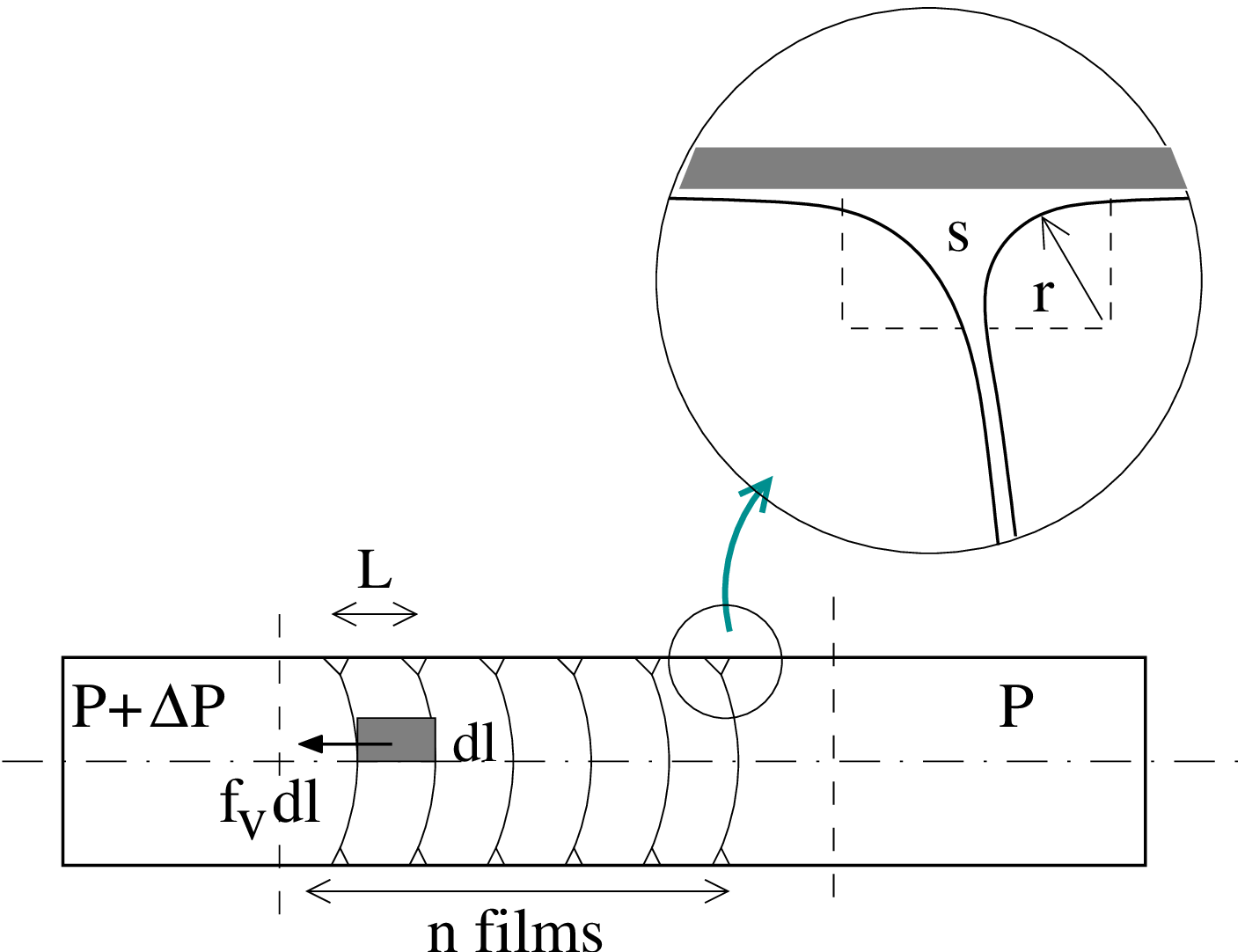} 
\end{minipage}
\hfill
\begin{minipage}{6.5cm}
\includegraphics[width=6.5cm]{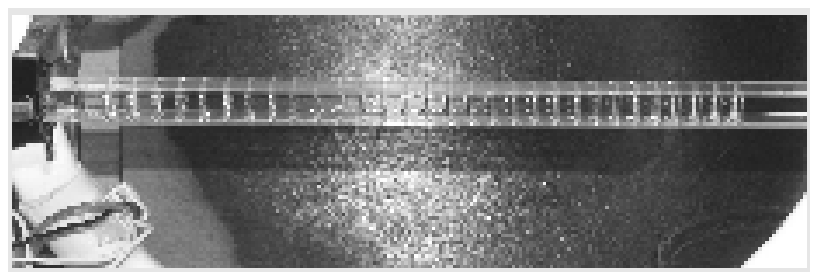}
\end{minipage}
\caption{(a) Schematic view of the system and definition of the parameters used in the text. The bubble train is moving to the right, pushed by a constant gas flux.  The pressure drop $\Delta P$ is balanced by the total viscous force exerted on the system  $\Omega$, limited by the two dashed lines. We define the viscous force $f_v$ so that the elementary area $L \,dl$, covering the interior of the tube and underlined in dark grey, is submitted to the viscous force $f_v dl$. The Plateau border cross section is magnified. 
(b) Bubble train in the square channel.
}
\label{system}
\end{figure}

Our foams are prepared from a $10\%$ in volume diluted commercial dish-washing fluid (Argos) in distilled water, leading to very fluid monolayers. We work on fresh prepared solutions in a room thermalized at $25^o$C. In these conditions, the dynamic viscosity, measured on a commercial viscosimeter (Schott Viscoclock) is $\eta$ = 1.04  10$^{-3}$ Pa.s,  the surface tension, measured with Wilhelmy plate, is $\gamma$= 30.2  10$^{-3}$ N.m$^{-1}$ and the solution density is $\rho$ = 996 kg m$^{-3}$. 
We use two different 30 cm long glass channels with respectively a $S$ = 1 cm$^2$ square cross section and a
$S$ = 0.12 cm$^2$ circular cross section. The bubble trains are prepared by air blowing in the foaming solution, directly in the tube. In order to control the geometry of the foam (bubble organization and size), a special attention is put on the air flow rate, on the position of the air injector with respect to the solution surface and on the tilting angle of the tube.
Once a regular bubble train is obtained, the solution is removed and the foam drains until the desired liquid amount is reached. All the films are perpendicular to the channel axis and the distance $L$ between two films, measured with a regular rule, is constant and of the order of few mm (see Fig. \ref{system}). The total amount of solution in the channel $V_{sol}$ is obtained by weighing. 
A thin liquid film wets the wall, but the largest part of the solution is trapped along the contact lines between the wall and the 
thin films separating the bubbles. The cross-section $s$ of these rims, called Plateau borders, is given by $s = V_{sol}/(n \,{\cal C})$, with $n$ the films number and ${\cal C}$ the channel perimeter. In the following, the control parameter used to characterize the amount of liquid in the foam is the Plateau border radius of curvature $r$ obtained from the geometrical relation $s = r^2 \,(2 - \pi/2)$, which assumes that the cross section is limited by the wall and by two quarters of circle, as depicted on Fig. \ref{system}. 
  The Plateau border radius, of the order of 1 mm, slightly varies between the 
top and the bottom of the horizontal tube and $r$ should be thus considered as its mean value. 
We only focused on dry foams, for which to successive Plateau borders do not touch each other, so $2 r < L$ (see Fig. \ref{system}).

Once prepared, the film train is pushed in the tube with a syringe pump (PHD 2000, Harvard Apparatus) at a constant velocity $V$ in the range $10^{-4} \, - \, 5 \, 10^{-3}$ m.s$^{-1}$. One channel end is at the atmospheric pressure, whereas the pressure at the other end is recorded by a pressure sensor (Keller). 
We performed several ways and back shifts of the foam within the tube, separated by a constant waiting time of few tens of seconds. Depending on the channel initial wetting, the first shift may exhibit an overpressure when the foam begins to move. Then, for each other shifts, a steady state is rapidly obtained, characterised by a constant value of the pressure, with a dispersion of the order of $5\%$. We counted the film number $n$ by hand and checked carefully that no rupture occured. 
During a shift, the Plateau border size seems to be homogeneous 
over the whole bubble train and constant in time. 

\section{Dimensional analysis}
The Reynolds number, based on the Plateau border size, may reach $Re = \rho r V/\eta \sim 10$. Nevertheless, the dynamical processes are governed by the quasi parallel hydrodynamical flow in the thin wetting film, for which inertial effects are negligible \cite{guyon}. We define the system $\Omega$ as the total amount of gas and solution in the bubble train (see Fig. \ref{system}). From the force balance on this system, we deduce  that 
the driving force $S \, \Delta P$, with $S$ the channel section and $\Delta P$ the total pressure drop, is exactly compensated by the viscous drag force between 
the wall and the bubble train. The tension surface forces, though very important in the problem, are internal 
forces and should not be taken into account in this force balance.
The dynamical behavior depends on the local pressure difference between air and liquid, but not on the mean 
pressure value. Each bubble plays thus a similar role and the system may be considered as periodic 
in the direction of the channel axis, with a period $L$, for all quantities except the mean pressure
that varies linearly from a periodic box to the other. Additionally, if the Plateau border size 
variation along the channel perimeter is negligible, the geometry is axisymmetric. The viscous force 
exerted by the wall on the elementary area $L dl$ between two films, drawn in grey on Fig.\ref{system},  is then simply $f_v dl$, with $f_v$ the viscous force per unit length of Plateau border given by
$
f_v = S \, \Delta P /(n {\cal C})
$.

This force depends on the foam velocity through the adimensional capillary number $Ca = \eta V / \gamma$. As shown in the following, it also depends on the Plateau border radius. If it decreases, the liquid phase is more confined, the velocity gradients increase and $f_v$ increases. Dimensionally, another length scale is thus needed
to built a length ratio involving $r$. The gravitational and inertial effects have been proved to be irrelevant. If the monolayers are governed by one of the two classical limiting cases of 
perfectly fluid or perfectly rigid interfaces, they do not introduce specific length scales either. Among geometrical lengths, the wetting film thickness $h$
is fixed by the dynamics itself and is not a control parameter.
The thin 
films across the channel are not involved in the dominating hydrodynamical processes, which are localised near the wall. So, as checked experimentally with various channel sections, the channel radius $R$ does not modify the viscous force. In cylindrical channel, this parameter may also 
appear through curvature effects, but the curvature $1/R$ is negligible in comparison with the Plateau border curvature. Any influence of the Plateau border radius should thus be expressed through the $r/L$ ratio, the distance $L$ between two films being the last remaining length scale. 
Finally, $f_v$ has the same dimension as $\gamma$ and the aim of our study is to determine experimentally $f_v/\gamma$ as a function of $Ca$ and of $r/L$, for a given foaming solution.

\section{Results}
\begin{figure}
\includegraphics[angle=-90,width=7.3cm]{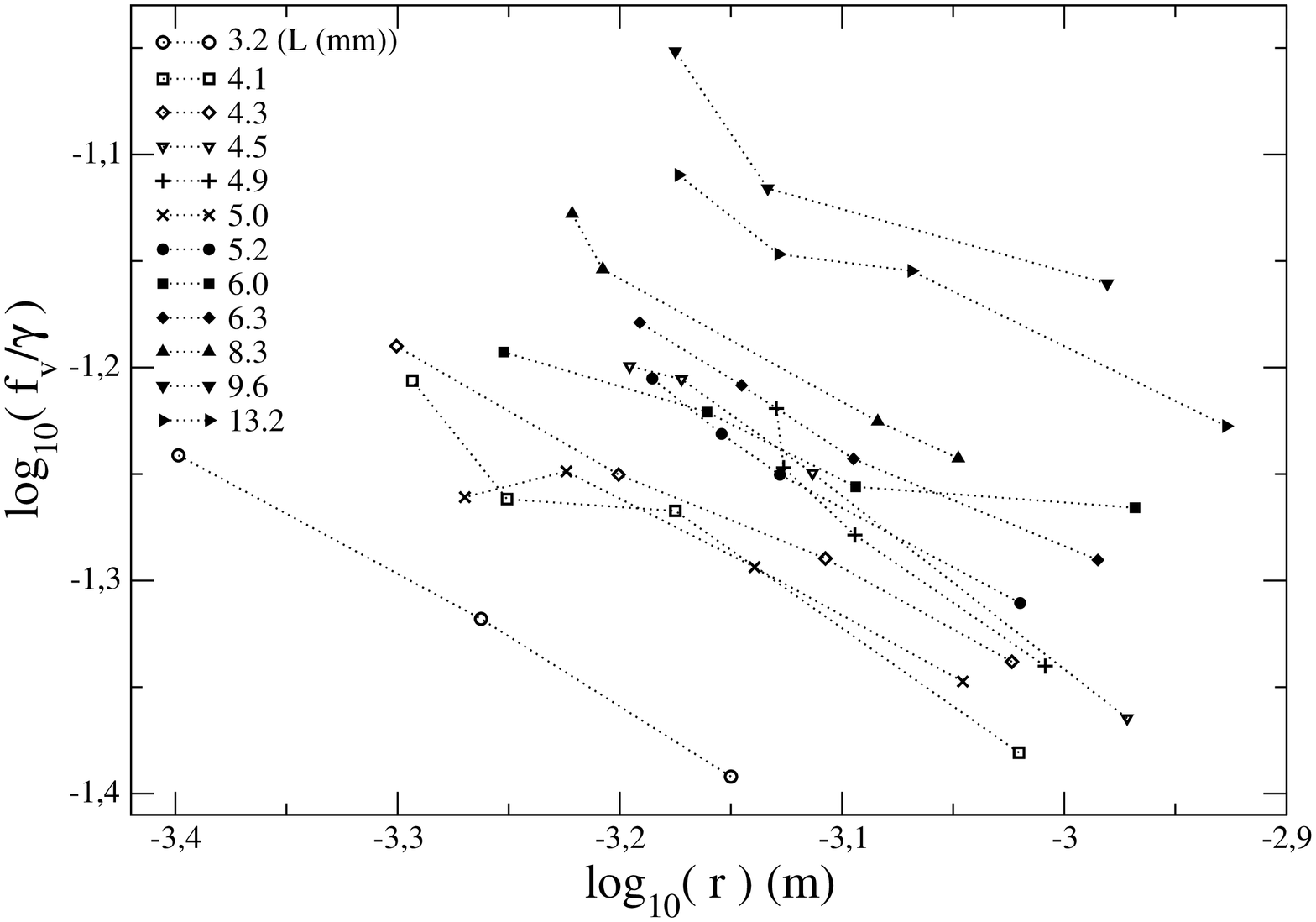}\hfill
\includegraphics[angle=-90,width=7.3cm]{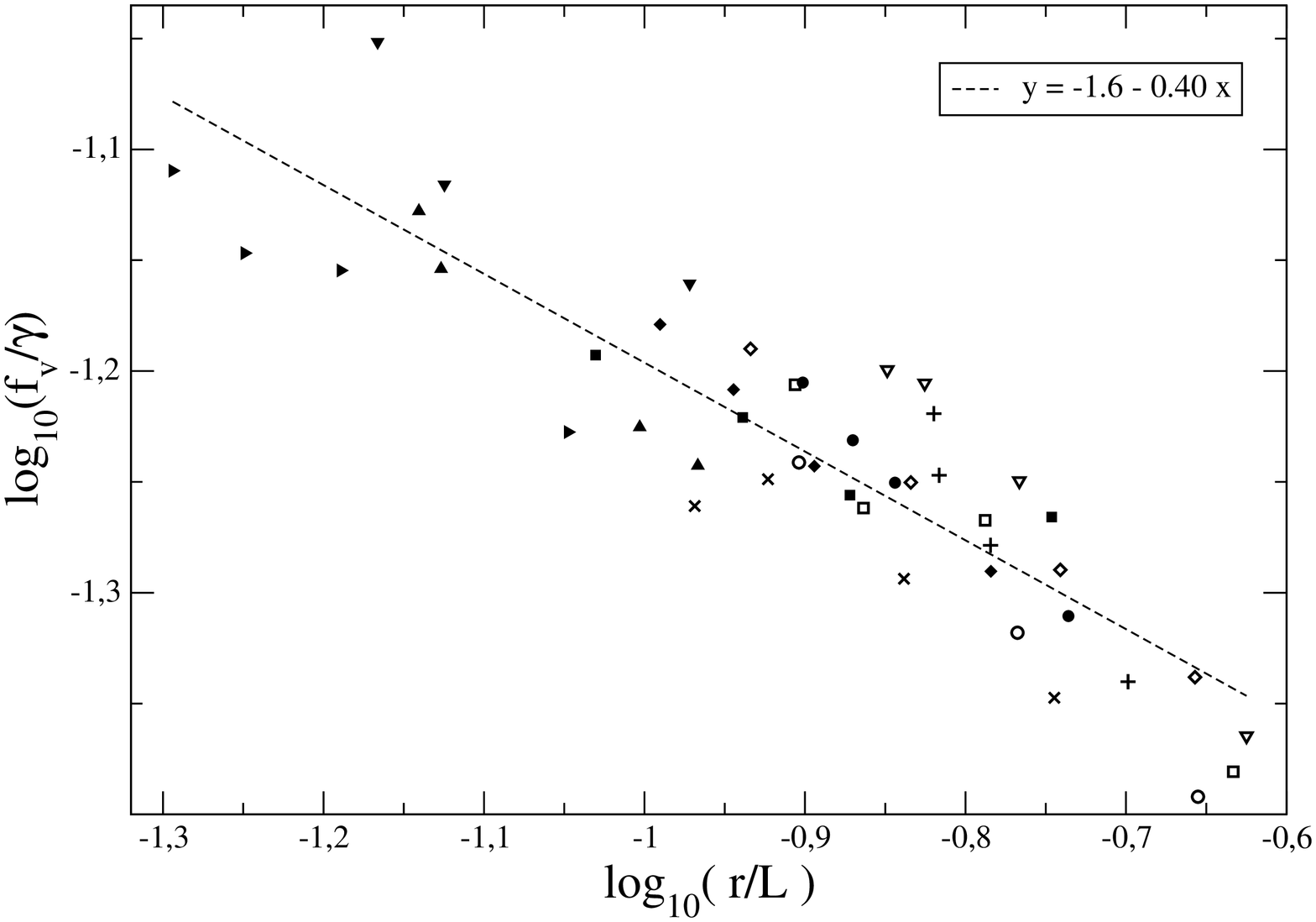}
\caption{(a) Viscous force per unit Plateau border length as a function of the Plateau border radius $r$, for several film spacings $L$. The capillary number $Ca= \eta V /\gamma$ is $10^{-4}$. The error bars are $\pm 5 \%$
 in both directions. (b) Same data as in graph (a), rescaled as a function of $r/L$. The dashed line is the best power fit, with an exponent $-0.4 \pm 0.05$.
}
\label{tubecarre}
\end{figure}

The first series of data, presented on Fig.\ref{tubecarre}(a), have been obtained in the square tube with a constant  velocity of $V$=3 mm.s$^{-1}$.  The film spacing was varied between 3 mm and 13 mm and, for each foam, the Plateau border radius was varied between 1.2mm and 0.4mm by drainage. 
The force decreases when the amount of solution increases or when film spacing 
 decreases, and is modified by a factor 2.5 on the parameter range investigated. This important variation with $L$ is in contradiction with the models based on a dissipation confined in the transition region
between the Plateau border and the thin film.
The dependence in the adimensional parameter $r/L$, predicted by the dimensional analysis, is 
confirmed by the good superposition of the various curves, once plotted as a function of $r/L$ (see Fig. \ref{tubecarre}(b)).
The noise is for a large part due to temperature fluctuations, that have been proved to modify the viscous force. They have been reduced in the experiments presented below. 

\begin{figure}[h]
\includegraphics[angle=-90,width=7.3cm]{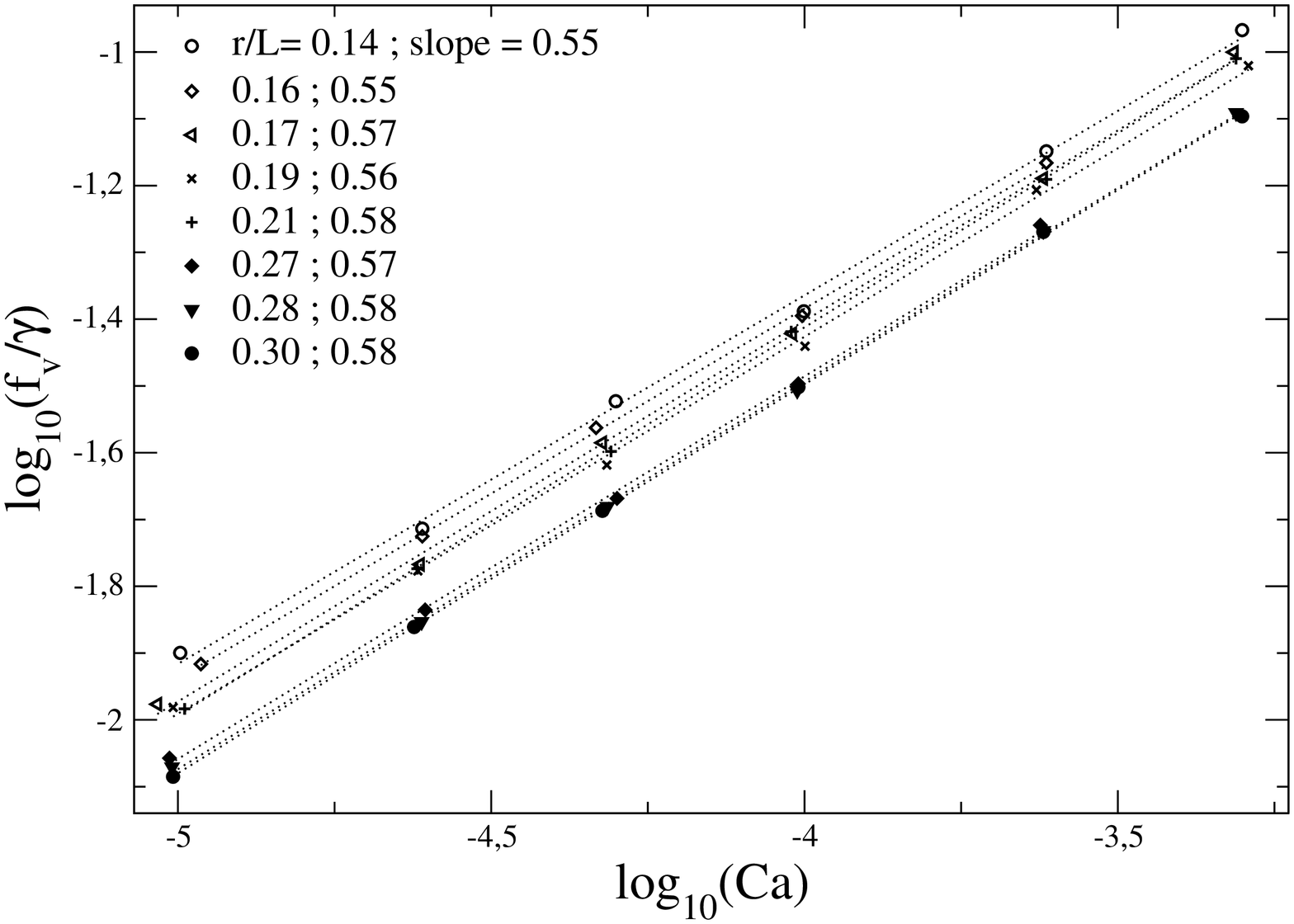}\hfill
\includegraphics[angle=-90,width=7.3cm]{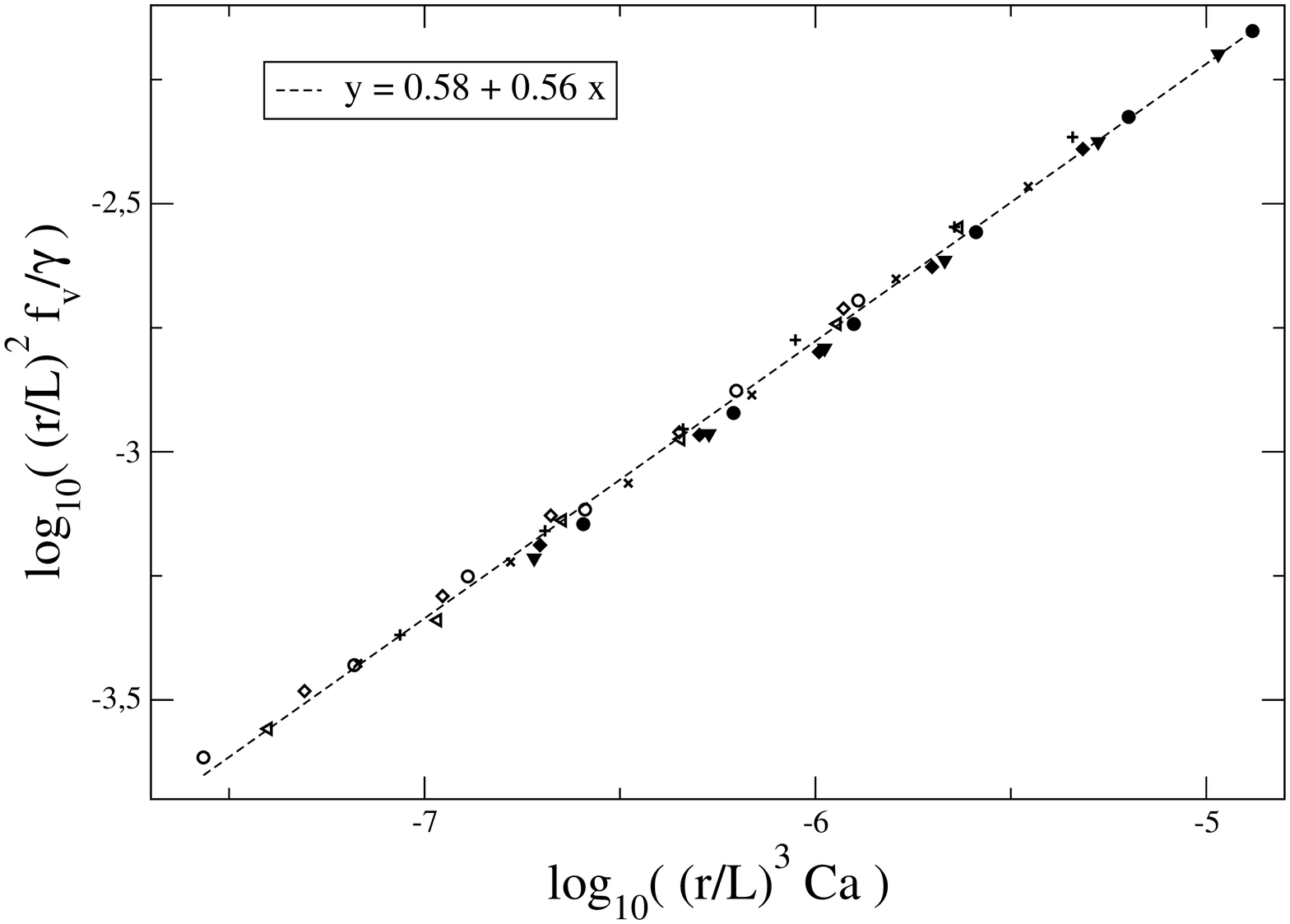}
\caption{(a) Viscous force per unit Plateau border length as a function of the capillary number $Ca= \eta V /\gamma$, for several film spacing and Plateau border radius. The dashed line are power law fit, with an exponent varying between 0.55 and 0.58. (b) Same data as in graph (a), rescaled according to equation \ref{force_scaling}. The rescaled data are well fitted by a power law with an exponent $0.56 \pm 0.02$.
}
\label{tuberond}
\end{figure}

The  velocity range is limited by the pressure sensor resolution at small velocity and by the film  breakage at high velocity. This parameter has been investigated in the small circular channel, which larger perimeter/section ratio allows to explore smaller velocities and to reach a velocity variation over 2 decades (see Fig. \ref{tuberond}(a)). The data, obtained for different values of $r$ and $L$, are well fitted by a power 
law with an exponent $0.56 \pm 0.02$. This exponent differs significantly from the 2/3 exponent predicted by Bretherton theory, initially established for a single bubble.
According to the theory detailed in the next section, the variations with the parameter $r/L$ and with the velocity are not independent. 
Once rescaled as proposed by eq. \ref{force_scaling}, discussed in the next paragraph, the dispersion of the experimental data falls from $\Delta f_v/f_v \sim 0.4$ on Fig. \ref{tuberond}(a) to the much smaller value  $\Delta f_v/f_v \sim 0.1$ on Fig. \ref{tuberond}(b), which confirms that the pertinent parameter governing the 
viscous force $f_v$ is  $\xi = (r/L)^3 Ca$.

\section{Modelisation}
The lubrication equation has been used to deal with the problem of bubble train motion in several previous 
papers, as recently by Denkov {\it et al} \cite{denkov05}. Nevertheless, despite the great simplification offered by this equation in contrast with the full Navier Stokes equation, 
it remains difficult to solve without additional assumptions. Our aim is not to determine an analytical expression for the viscous force, but simply to show, by a pertinent rescaling, that this force can be expressed as a function of a single adimensional parameter only, namely $\xi= (r/L)^3 Ca$. 
\begin{figure}[h]
\centerline{
\includegraphics[width=6.5cm]{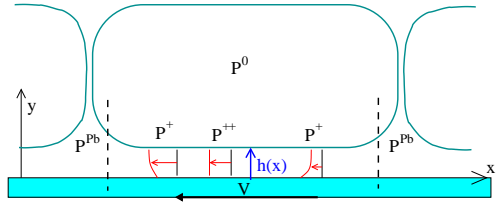}
}
\caption{Schematic representation of the velocity and pressure distribution in the film. The pressure in the thin film is not determined in this paper, but, by analogy with simpler geometries, we can say that its value $P^{+}(x)$ is larger than the reference
value $P^{Pb}$ in the Plateau borders \cite{guyon}. This allows to give an idea of the velocity profiles in the bubble frame on both sides of the maximum pressure value $P^{++}$.  The system $d\omega$ used in the
text is limited by the dashed lines, with a thickness dl in the transverse direction. 
}
\label{lubrif}
\end{figure}

The hydrodynamical problem is solved in the bubble train 
frame for the wetting film between two successive Plateau borders, defined by its thickness $h(x)$.  We assume a plane wall and a translational invariance in the direction $z$, normal to the wall velocity  $-V {\bf u}_x$ and to the wall normal  ${\bf u}_y$. This open system, called $d\omega$ and depicted on Fig. \ref{lubrif}, is limited by the abscissa $x=0$ and $x=L$, where the film has the same thickness $h(0)=h(L)$ and has the curvature of the equilibrium Plateau border $\partial^2 h/\partial x^2 (0 \mbox{ or } L)=1/r$ (see Fig. \ref{lubrif}). The local pressure variations in the film are much larger than  the pressure drop between two successive Plateau borders, which is neglected.
Classically, the lubrication approximation leads to the following expression
$ \eta \,  \partial^2v_x / \partial y^2= \partial P / \partial x$  \cite{guyon}. 
With the boundary conditions $v(x,y=0)= -V$ at the wall (by convention $V>0$) and a vanishing tangential force at the liquid/air interface at the position $y=h(x)$ (fluid interface), 
 we obtain the parabolic velocity field sketched on Fig.\ref{lubrif}
\begin{equation}
 v_x= {1 \over 2\eta} \drp{P}{x} y(y-2h(x))-V 
\label{vx}
\end{equation} 
By integration of eq.\ref{vx}, we get a relation between the film profile and $Q$, the  total flux of solution in the thin wetting film, in the bubble train frame ($Q<0$ with our conventions). Finally, using the Laplace condition $P= - \gamma \partial^2 h /\partial x^2$ to suppress the unknown parameter P, we get the equation governing the film profile :
\begin{equation}
\drp{^3h}{x^3}= -{1\over \gamma} \drp{P}{x} = {3 Ca \over h^2} \left(1+ {Q \over V h} \right)
\label{eqenh}
\end{equation} 
In contrast with previous studies on single long bubbles in capillaries, the value of Q can not be determined
from boundary conditions far from the bubble front and a new relation must be taken into account to close the problem. 
An integral condition arises from the global force and energy balances. The force exerted by the fluid on the wall is given by
\begin{equation}
{f_v \over \gamma} = {\eta \over \gamma} \int_0^L \drp{v_x}{y}(y=0) dx = \int_0^L {3 Ca \over h} \left(1+ {Q \over V h} \right)dx
\label{force}
\end{equation} 
The power injected into the system is entirely dissipated in the viscous modes. By periodicity in the x direction, there is no net energy flux out of the considered open system $d\omega$, if we neglect the dissipation in the Plateau border itself. As the only moving boundary is the wall, the injected power is thus simply $f_v V$ and the energy balance is
\begin{equation}
{f_v V \over \gamma} = {\eta \over \gamma} \int_0^L\int_0^{h(x)} \left (\drp{v_x}{y} \right )^2 dx dy =  V \int_0^L {3 Ca \over h}
\left(1+ {Q \over V h} \right)^2 dx
\label{energie}
\end{equation} 
Once developed as polynomials of the variable Q, eqs.\ref{force},\ref{energie} lead to 
$Q/ V  =-  \int_0^L h^{-2}dx /  \int_0^L  h^{-3}dx $.

The only way to suppress all the physical parameters from eq.\ref{eqenh} is to use the rescaling  
$\bar{h} = h /(L Ca^{1/3})$ and $\bar{x}= x/L$, leading to 
\begin{equation}
\drp{^3\bar{h}}{\bar{x}^3}= {3 \over \bar{h}^2} \left(1+ {\bar{Q} \over \bar{h} } \right ) \; \mbox{  with   } \; \bar{Q} = {Q \over V L Ca^{1/3}} = - {\int_0^1 \bar{h}^{-2}\bar{dx} \over \int_0^1 \bar{h}^{-3}\bar{dx}} 
\end{equation}
The solution $\bar{h}$  of this integro-differential equation is entirely determined by the three
boundary conditions $\bar{h}(0)=\bar{h}(1)$ and 
$\partial^2\bar{h}/\partial \bar{x}^2(0 \mbox{ and } 1)= \xi^{-1/3}$, which only depend on the parameter 
$\xi =Ca (r/L)^3$. 
Finally, the viscous force given by eq. \ref{force} can be expressed as
\begin{equation}
f_v  = 3 \gamma Ca^{2/3} \int_0^1 \bar{h}^{-1}  \left(1+ {\bar{Q} \over \bar{h} } \right )\bar{dx} = \gamma (r/L)^{-2} \, F((r/L)^3 Ca)
\label{force_scaling}
\end{equation}
The integral in eq.\ref{force_scaling} is a function of $\bar{h}$ and thus only depends on $\xi$. This leads to the second part of eq.\ref{force_scaling}  which was successfully used
to rescale the experimental data on Fig.\ref{tuberond}. From these calculations we cannot deduce any specific form for the function $F$. The power law fit proposed on Fig.\ref{tuberond} is thus only phenomenological.
The case of a rigid interface, leading to the condition $v(x,y=h)= 0$, only induces variations of some numerical parameters
in the equations of motion and allows to derive the general expression for the viscous force, with an other unknown function $F$.

\section{Discussion}
The phenomenological scaling law obtained from our experimental data in the cylindrical tube is (see Fig. \ref{tuberond})
\begin{equation}
f_v=(3.8 \pm 0.2)  \, \gamma\, \left(r \over L \right)^{-2} \left(\left(r \over L \right)^3 Ca\right) ^{0.56 \pm 0.02}
\label{fv_exp}
\end{equation}
Replacing the capillary number in eq.\ref{fv_exp} by its value in the first series of experiments in the square tube, $Ca =10^{-4}$, we get $f_v \sim 2.2 \, 10^{-2} \,\gamma (r/L)^{-0.32 \pm 0.06}$,  consistently with the scaling $f_v \sim 2.5 \, 10^{-2} \, \gamma(r/L)^{-0.40 \pm 0.05}$ obtained from a direct fit of these experimental data  (see Fig. \ref{tubecarre}). The channel shape and size have thus a negligible influence. 
Both series of data are finally in fair agreement with the theoretical prediction eq.\ref{force_scaling} with the unknown function $F(\xi) = (3.8 \pm 0.2)\,  \xi^{\,0.56 \pm 0.02}$.

The effective power law obtained for $F$ probably depends on the parameter range for $\xi$, which should be investigated in further experiments.
As the parameter $r$ is not defined in single bubble experiments and was usually not measured in studies involving foams or bubble trains, quantitative comparisons with previous results are difficult to performed. 
With our previous experimental set-up \cite{cantat04}, the liquid fraction  was not controlled, but the ratio $r/L \sim 0.05$ was roughly estimated on the images of the foam films. The obtained power law  was 2/3, excepted for capillary numbers smaller than 
$Ca \sim 6 \, 10^{-5}$ (or $\xi \sim 7\, 10^{-9}$), for which a smaller exponent was found.  For the motion of isolated bubbles in capillaries, without surfactant, Schwartz {\it et al} also found different power laws (for the film thickness $h$, which scaling is related to the $f_v$ scaling), depending on the bubble aspect ratio \cite{schwartz86}. In this case, the Plateau border radius of curvature should be replaced by the channel radius in the modelisation. The power law $2/3$ is obtained for $\xi > 10^{-8}$, whereas $1/2$ is found for smaller values.  The exponents obtained on Fig. \ref{tuberond}(a) also slightly  increase with increasing $\xi$, but the values for $\xi$ are larger than $10^{-8}$.  

Finally, the influence of the interfacial rheology on the viscous force has not been studied in this paper, but the 
general expression eq.\ref{force_scaling}, with appropriate values for $F$, is compatible with previous results and provides a more general frame of interpretation.
With $F(\xi) \sim \xi^{2/3}$ the Bretherton exponent 2/3 is recovered for the velocity. In this case, $f_v$ should be independent on the liquid fraction, which was already pointed out by Schwartz and Princen \cite{schwartz87}. On the other hand, the function $F(\xi) \sim \xi^{1/2} $ leads to 
the law $f_v \sim \gamma (L Ca/ r)^{1/2}$, obtained theoretically for bubbles with rigid interfaces by Denkov {\it et al} \cite{denkov05} and for soft particles in similar geometry \cite{meeker04}.

\section{Conclusion}

In this paper, we present experimental results measuring the influence of the foam liquid fraction on the viscous force exerted at a solid boundary. We derive an original scaling law for this force, based on simple arguments,  which provides a
new frame to understand the various predictions found in the literature for this very controversial problem. 
We especially underline the strong coupling between the power law obtained for the velocity 
and the dependence on the liquid fraction. 

\section{Acknowledgements}
The authors thank R. Delannay and G. Le Ca\"er for many constructive discussions. We are very grateful to N. Denkov for making his work available prior to publication. We thank the R\'egion Bretagne for its financial support.

\end{document}